\documentclass{elsart}

\usepackage{graphicx}
\usepackage{amssymb}

\usepackage{supertabular}

\begin{document}

\begin{frontmatter}

\title{Measurements of atmospheric muon spectra at mountain altitude}

\author[tok]{T. Sanuki\corauthref{cor1}},
\corauth[cor1]{Corresponding author.}
\ead{sanuki@phys.s.u-tokyo.ac.jp}
\author[tok]{M. Fujikawa},
\author[tok]{K. Abe},
\author[tok]{K. Anraku\thanksref{kanagawa}},
\author[tok]{Y. Asaoka\thanksref{icrr}},
\author[tok]{H. Fuke},
\author[tok]{S. Haino},
\author[tok]{M. Imori},
\author[tok]{K. Izumi},
\author[kob]{T. Maeno},
\author[kek]{Y. Makida},
\author[tok]{N. Matsui},
\author[tok]{H. Matsumoto},
\author[tok]{H. Matsunaga\thanksref{tuku}},
\author[tok]{M. Motoki\thanksref{toho}},
\author[tok]{J. Nishimura},
\author[kob]{M. Nozaki},
\author[tok]{S. Orito\thanksref{ori}},
\author[kek]{M. Sasaki\thanksref{nasa}},
\author[kob]{Y. Shikaze},
\author[tok]{T. Sonoda},
\author[kek]{J. Suzuki},
\author[kek]{K. Tanaka},
\author[kob]{Y. Toki},
\author[kek]{A. Yamamoto},
\author[tok]{Y. Yamamoto},
\author[kob]{K. Yamato},
\author[kek]{T. Yoshida},
\author[kek]{K. Yoshimura}

\address[tok]{The University of Tokyo,
 Bunkyo, Tokyo 113-0033, Japan}
\address[kob]{Kobe University,
 Kobe, Hyogo 657-8501, Japan}
\address[kek]{High Energy Accelerator Research Organization (KEK),
 Tsukuba, Ibaraki 305-0801, Japan}

\thanks[kanagawa]{Present address: Kanagawa University,
 Yokohama, Kanagawa 221-8686, Japan}
\thanks[icrr]{Present address: ICRR, The University of Tokyo,
 Kashiwa, Chiba 227-8582, Japan}
\thanks[tuku]{Present address: University of Tsukuba,
 Tsukuba, Ibaraki 305-8571, Japan}
\thanks[toho]{Present address: Tohoku University,
 Sendai, Miyagi 980-8578, Japan}
\thanks[ori]{deceased.}
\thanks[nasa]{Present address:
 National Aeronautics and Space Administration,
 Goddard Space Flight Center, Greenbelt, MD 20771, USA;
 NAS/NRC Research Associate.}

\begin{abstract}
We report new measurements of the atmospheric muons
 at mountain altitude. 
The measurement was carried out with the BESS detector
 at the top of Mt. Norikura, Japan. 
The altitude is 2,770~m above sea level. 
Comparing our results with predictions given by some interaction models,
 a further appropriate model has been investigated. 
It would improve
 accuracy of atmospheric neutrino calculations. 
\end{abstract}

\end{frontmatter}

\clearpage

\section{Introduction} \label{sec:intro}

Atmospheric muons observed in the deep atmosphere are carrying
 information on the interaction
 of primary and secondary cosmic-ray particles in the atmosphere. 
Since decay probability and energy loss rate depend on energy,
 high-statistics measurement of spectral shape
 can provide useful knowledge
 to understand propagation process of cosmic-ray particles
 through the atmosphere. 
For a more detailed study of neutrino oscillation phenomena
 observed in atmospheric neutrinos
 it is essentially important to reduce the systematic errors
 in the predicted energy spectra of neutrinos.
In order to improve the accuracy of the predictions,
 a detailed understanding of hadronic interactions,
 as well as precise information of primary cosmic-ray fluxes,
 is indispensable. 
As for absolute fluxes of cosmic-ray protons and helium nuclei
 below 100~GeV,
 a very precise measurement was carried out
 by using the BESS detector \cite{bessphe}. 
Two independent experiments,
 the AMS \cite{amsp,amshe} and the BESS,
 show extremely good agreement with each others
 in their results. 
Since production and decay process of muons are accompanied
 by neutrino productions,
 the muons fluxes correlate directly to the hadronic interactions. 
Thus
 the precise measurement of atmospheric muon spectra can also provide
 important information
 to improve accuracy of atmospheric neutrino calculations. 

We measured absolute fluxes of atmospheric muons
 at the top of Mt. Norikura, Japan
 with the BESS detector,
 which was the same apparatus as we utilized to
 measure sea-level muons \cite{motokimu}
 as well as the primary protons and helium nuclei \cite{bessphe}. 
These measurements provide
 fundamental and very useful information
 about hadronic interactions and geomagnetic effects. 

\section{BESS experiment} \label{sec:bess}

\subsection{BESS Detector}
\begin{figure}
\includegraphics[width=\textwidth]{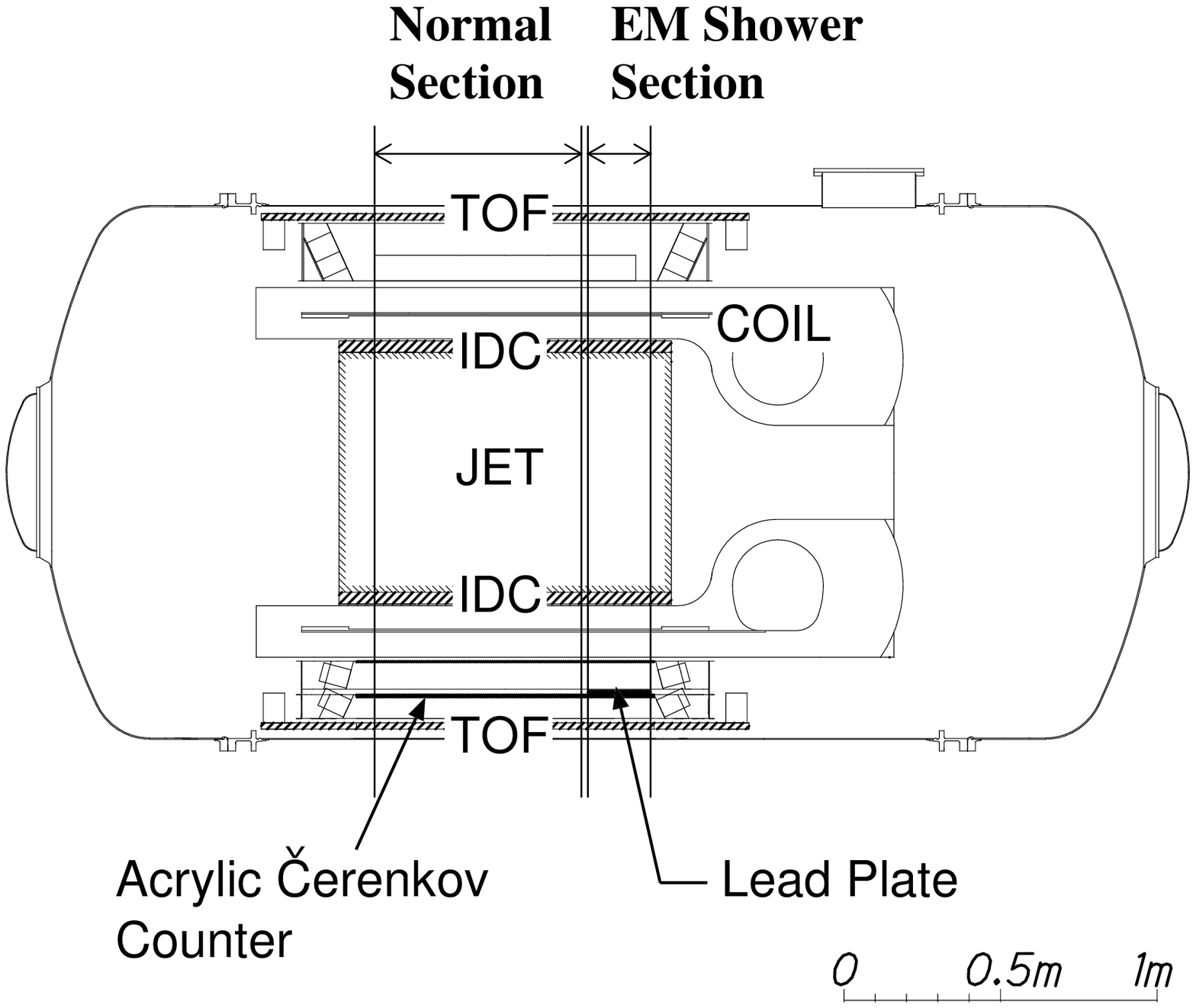}
\caption{Schematic cross-sectional view of the BESS instrument.}
\label{fig:besscross}
\end{figure}

The BESS ($\underline{\rm B}$alloon-borne $\underline{\rm E}$xperiment
 with a $\underline{\rm S}$uperconducting 
 $\underline{\rm S}$pectrometer)
 detector \cite{orito,yamamoto1994,agel,detector,newtof} is
 a high-resolution spectrometer with a large acceptance
 to perform highly sensitive searches
 for rare cosmic-ray components,
 and precise measurement of the absolute fluxes
 of various cosmic rays. 
Fig.~\ref{fig:besscross} shows a schematic cross-sectional view
 of the BESS instrument.
In the central region,
 a uniform magnetic field of 1 Tesla is provided
 by using a thin superconducting solenoidal coil. 
A magnetic-rigidity ($R \equiv Pc/Ze$) of
 an incoming charged particle is measured
 by a tracking system,
 which consists of a jet-type drift chamber (JET)
 and two inner-drift-chambers (IDC's)
 inside the magnetic field. 
The deflection ($R^{-1}$) and its error
 are calculated
 for each event by applying a circular fitting
 using up-to 28 hit points,
 each with a spatial resolution of 200~$\mu$m. 
The maximum detectable rigidity (MDR) was estimated to be 200~GV. 
Time-of-flight (TOF) hodoscopes provide the velocity ($\beta$)
 and energy loss ($\d E/\d x$) measurements. 
A 1/$\beta$ resolution of 1.6~\% was achieved in this experiment. 

Particle identification was made
 by requiring proper 1/$\beta$ as well as $\d E/\d x$
 as a function of the rigidity. 
An electromagnetic shower counter has been equipped
 for electron/muon separation. 
It consists of a 2 $X_0$ thick lead plate and
 an acrylic ${\rm \check{C}erenkov}$ counter. 
The lead plate covers about 1/5 of the total acceptance. 

The simple cylindrical shape and the uniform magnetic field
 make it simple and reliable
 to determine the geometrical acceptance precisely. 
The live data-taking time was measured exactly
 by counting 1~MHz clock pulses with a scaler system
 gated by a ``ready'' status that controls the first-level trigger. 
The resultant live-time ratio was as high as 98.8~\%
 through the measurement. 

\subsection{Data samples}
\begin{figure}
\includegraphics[width=\textwidth]{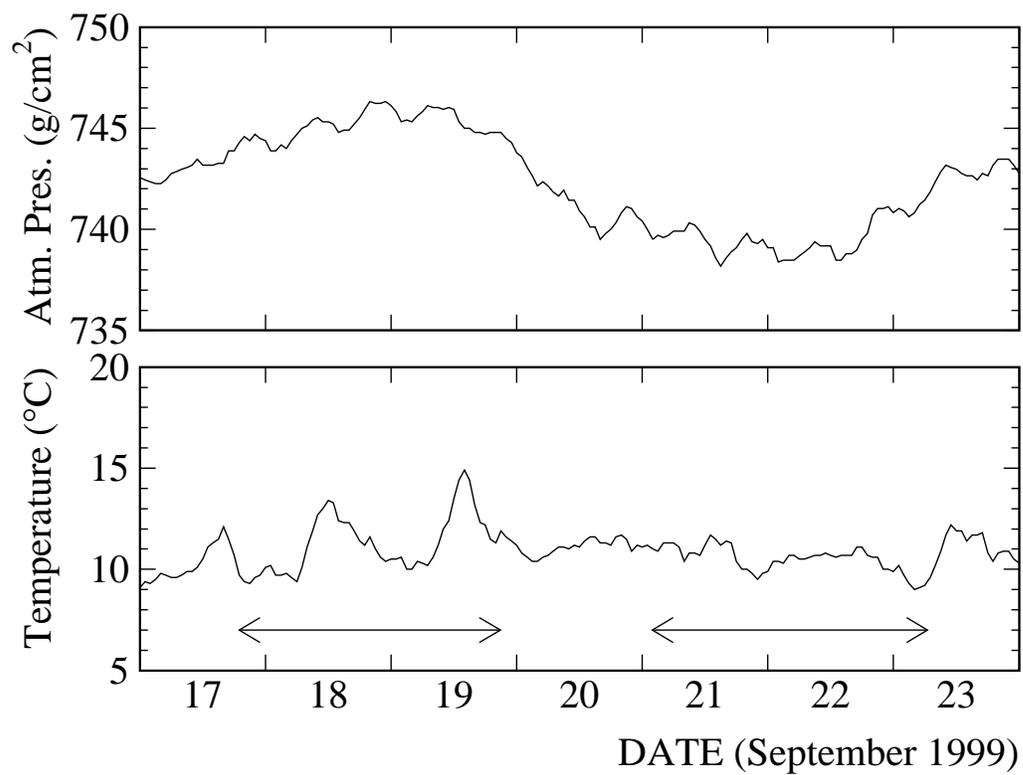}
\caption{Atmospheric pressure and temperature during the observation.
 The experiment was performed
 during two periods of 17th -- 19th and 21st -- 23th.}
\label{fig:env}
\end{figure}

The atmospheric cosmic-ray events were observed at Norikura Observatory,
 ICRR, the University of Tokyo, Japan. 
It is located at $36^\circ~06'N, 137^\circ~33'E$,
 which is about 40~km southeast of Kamioka. 
The altitude is 2,770~m above sea level. 
The vertical geomagnetic cutoff rigidity is 11.2~GV \cite{shea}. 
The experiment was performed
 during two periods of 17th -- 19th and 21st -- 23th of September 1999. 
During the observation, the atmospheric pressure and temperature
 varied as shown in Fig.~\ref{fig:env}. 
The mean (root-mean-square) atmospheric pressure and temperature were
 742.4 (2.9)~${\mathrm {g/cm^2}}$ and 10.9 (1.1)~${\mathrm {^\circ C}}$,
 respectively. 
The total live data-taking time was about 4 days. 

\section{Data analysis} \label{sec:analysis}

\subsection{Data Reduction}
In this analysis, the BESS detector was divided into two sections
 as shown in Fig.~\ref{fig:besscross}; i.e.,
 the ``normal'' section and the ``EM shower'' section. 
The EM shower section includes the shower counter. 
The normal section was used to measure muon momentum spectra. 
In order to estimate small contaminations of electrons,
 the data samples recorded in the EM shower section were examined. 

In the first stage of data reduction,
 we selected events with a single-track
 fully contained within a fiducial volume. 
The fiducial volume was defined
 as a central region in the JET chamber. 
This definition
 ensured the longest track
 to achieve the highest resolution in the rigidity measurement. 
The zenith angle ($\theta _z$) was limited
 to be $\cos \theta _z \geq 0.98$ to obtain near vertical fluxes. 
The single-track event was defined
 as an event which has
 only one isolated track
 and one or two hit-counters in each layer of the TOF hodoscopes. 
Since the major component of the observed particles was muons,
 those which can easily penetrate the detector,
 almost all particles passed this selection criterion. 
The single-track selection eliminated very rare interacting events. 
In order to verify the selection,
 events were scanned randomly. 
It was confirmed that
 1,995 out of 2,000 visually-identified single-track events
 passed this selection criteria
 and interacting events were fully eliminated. 
Thus, the track reconstruction efficiency was determined
 to be $99.8 \pm 0.1$~\% for the single-track event. 
The single-track events selected in this stage
 were followed by the succeeding analysis. 

\begin{figure}
\includegraphics[width=\textwidth]{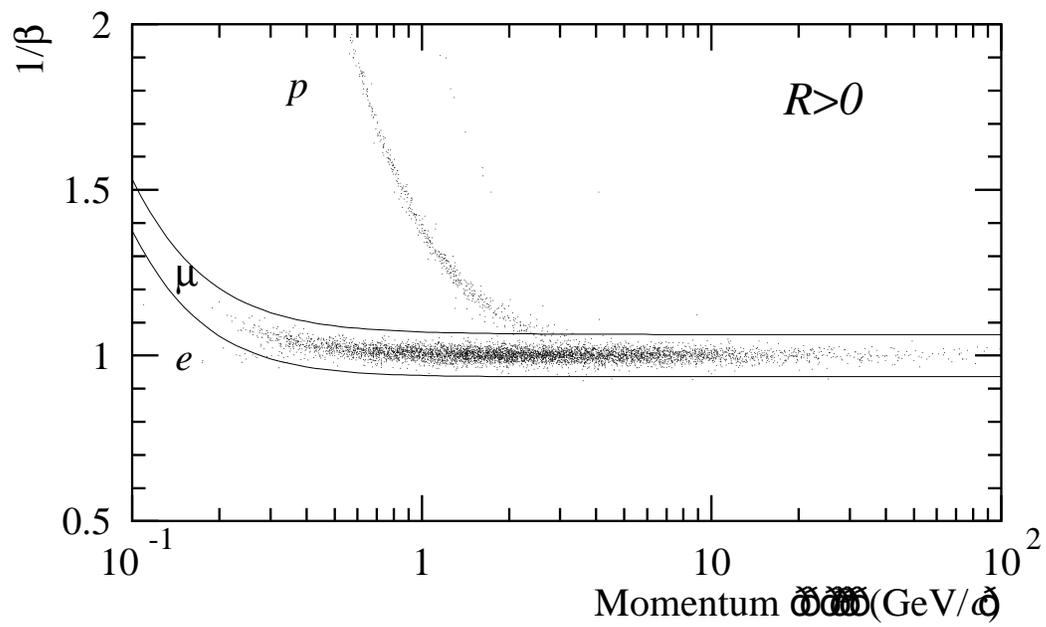}
\caption{Positive muon selection in 1/$\beta$ vs. momentum.
 Muons were selected with a high efficiency.
 Proton and positron contaminations were subtracted.}
\label{fig:mu+select}
\end{figure}

\begin{figure}
\includegraphics[width=\textwidth]{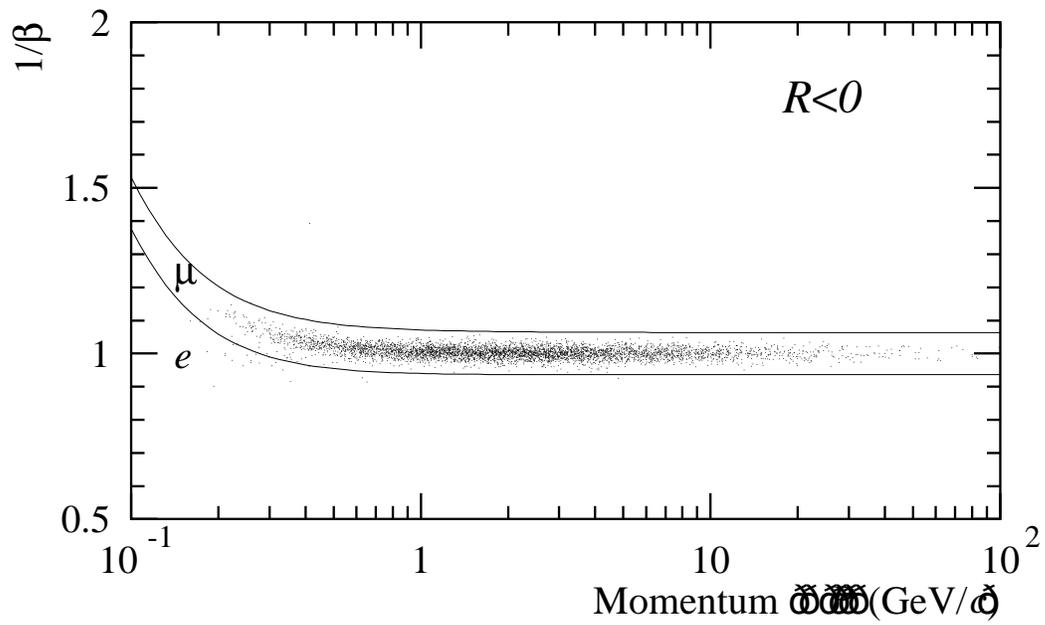}
\caption{Negative muon selection in 1/$\beta$ vs. momentum.
 Muons were selected with a high efficiency.
 Electron contamination was subtracted.}
\label{fig:mu-select}
\end{figure}

The $\d E/\d x$ inside the TOF hodoscope was examined
 to ensure that the particle was singly charged. 
The particle identification was performed
 by requiring proper 1/$\beta$ as a function of the momentum
 as shown in Figs.~\ref{fig:mu+select} and \ref{fig:mu-select}. 
Since the 1/$\beta$ distribution was well described by Gaussian and
 a half width of the 1/$\beta$ selection band was set at 3.89 $\sigma$,
 the efficiency was very close to unity. 
Taking into account a small deviation from pure Gaussian,
 the efficiency was evaluated to be 99.9 \%. 
In order to assure accuracy of the momentum measurement,
 event qualities such as $\chi^2$ in the track fitting procedure
 were required. 
The efficiency of this quality-check was evaluated
 to be $96.3 \pm 0.1$~\%
 as a ratio of the number of good quality events to all events
 between 10 and 20~GeV/$c$. 
Since the quality-check required consistency
 between the fitted track and each hit in the tracking detector,
 its efficiency does not depend on the momentum
 in higher momentum region
 where multiple scattering can be neglected. 
Accordingly the above value of the quality-check efficiency was applied 
 to higher momentum region. 
Before and after the quality-check
 and succeeding correction of its efficiency,
 the obtained flux above 10~GeV/$c$
 changed no more than $\pm 2.2$~\%. 
These changes were treated as systematic errors in the quality-check. 
Below 10~GeV/$c$, quality-check was not imposed,
 because momentum resolution before the quality-check was high enough
 to measure the momentum. 
Thus there was no systematic error associated with the quality-check. 

With the data reduction described above,
 707,563 muon candidates were selected. 
The combined efficiency in the data reduction process was
 $99.6 \pm 0.1$~\% and $95.8 \pm 2.2$~\% below and above 10~GeV/$c$,
 respectively. 

\subsection{Contamination estimation}
\begin{figure}
\includegraphics[width=\textwidth]{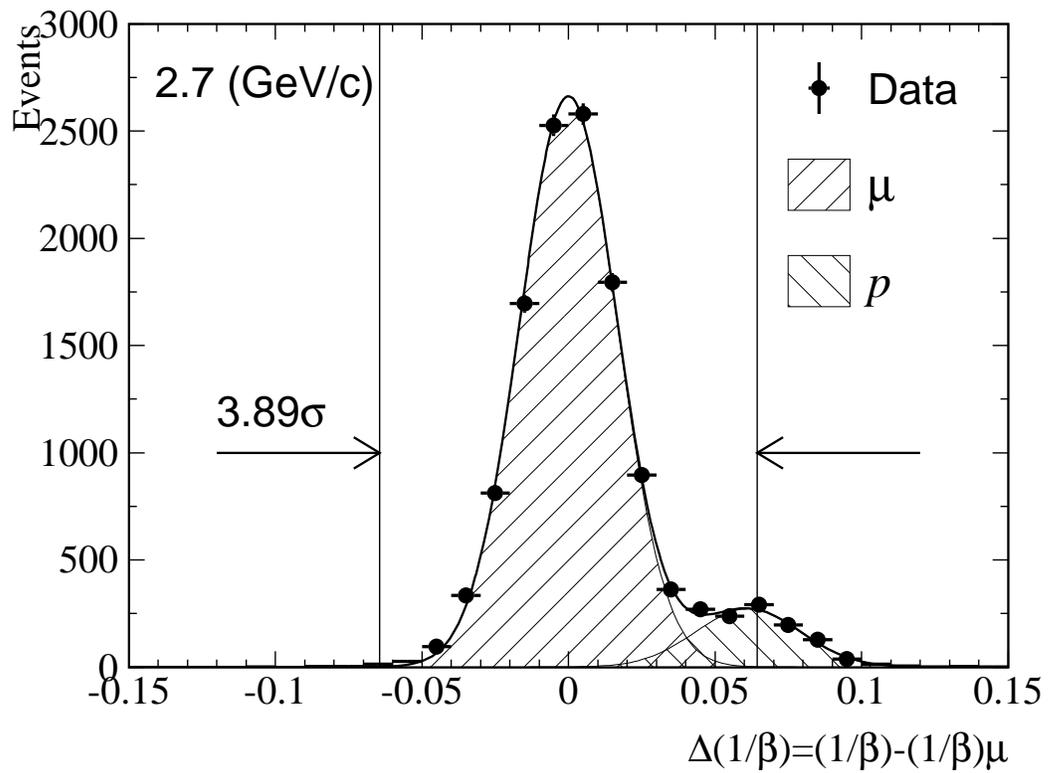}
\caption{An estimation of proton contamination by TOF information.
 Muons have a peak at $\Delta(1/\beta)=0$.
 Filled circles are data at around 2.7~GeV/$c$.
 Best fitting curves are also shown.}
\label{fig:pfrac}
\end{figure}

As seen in Figs.~\ref{fig:mu+select} and \ref{fig:mu-select},
 main background sources were protons for positively charged muons
 whereas electrons and positrons
 for both positively and negatively charged muons. 
Protons were identified by examining 1/$\beta$ distribution. 
Below 1.6~GeV/$c$, protons were clearly separated from muons. 
Between 1.6 and 3.2~GeV/$c$,
 the proton contamination in the muon candidates
 and its error were estimated
 by fitting the 1/$\beta$ distribution with a double-Gaussian function
 as shown in Fig.~\ref{fig:pfrac}. 
The ratio of observed protons to muons was found to be well described
 as a simple power law as $p/\mu^+ = 0.24 \cdot P^{-0.80}$
 in a momentum region of 1 -- 3~GeV/$c$. 
The proton contamination above 3~GeV/$c$ was estimated
 by extrapolating this power law. 
Above around 10~GeV/$c$,
 however, 
 the $p/\mu^+$ ratio will be constant. 
It is reasonably assumed that
 pions are produced
 having a similar spectral index of the parent protons
 and then decay to muons,
 those which can reach the mountain altitude before they decay
 in a higher momentum region. 
Under this assumption, 
 the spectral index of muons is similar to that of the parent protons,
 resulting in a constant $p/\mu^+$ ratio. 
This tendency was well reproduced
 by two Monte Carlo simulations \cite{cosmos};
 one simulation incorporated the FRITIOF7 \cite{fritiof7}
 as a hadronic interaction model and
 the other was developed
 based on the DPMJET-III event generator\cite{dpmjet3}.
In this analysis, the $p/\mu^+$ ratio above 10~GeV/$c$ was assumed
 to take a constant value of 3.8\%. 
These proton contaminations were subtracted from the muon candidates. 
An error originated from the proton subtraction
 between 1.6 and 3.2~GeV/$c$ was estimated
 during the fitting procedure to be less than 1~\%. 
In a higher momentum region,
 the $p/\mu^+$ ratios predicted by the two Monte Carlo simulations
 agreed with each other within $\pm 1\%$. 
These discrepancies were included into systematic errors. 

The muon candidates might include pion contamination,
 since it has the same signature and is hardly identified. 
According to the same Monte Carlo simulations
 that was utilized for $p/\mu^+$ ratio estimations,
 $\pi/\mu$ ratio was estimated to be less than 1~\% below 40~GeV/$c$
 and less than 3~\% between 40 and 100~GeV/$c$. 
It is smaller than the statistical errors
 over the whole momentum range in this analysis. 
The pions were neglected in this analysis both
 in contamination subtraction and systematic error estimation. 

\begin{figure}
\includegraphics[width=\textwidth]{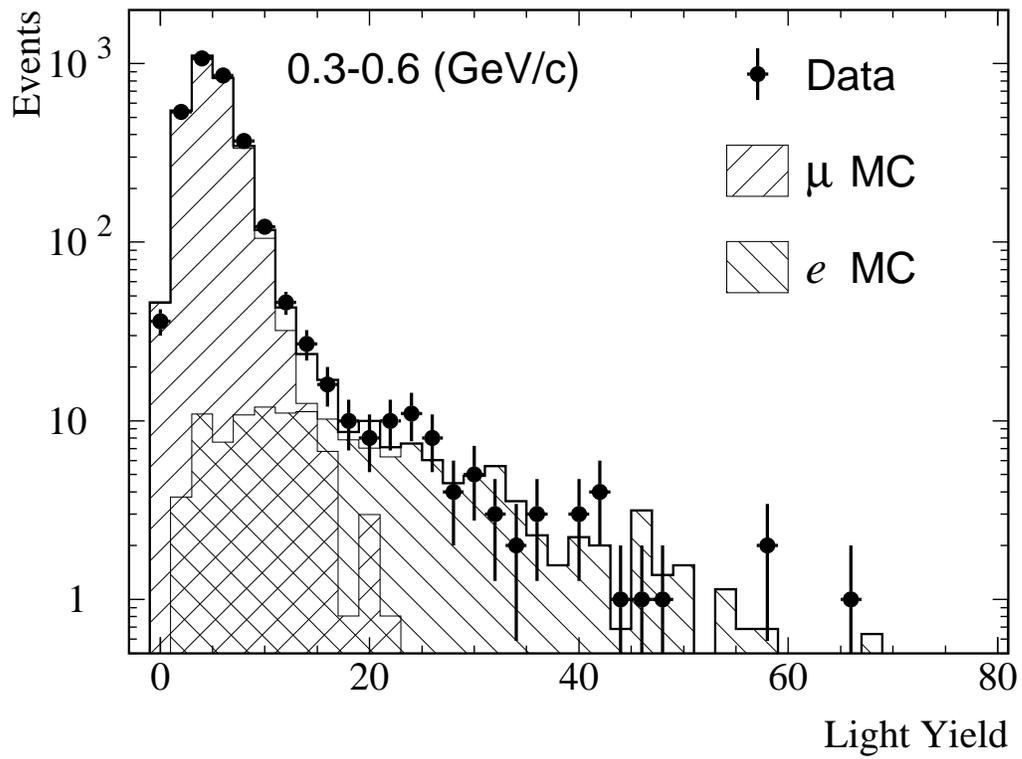}
\caption{An estimation of electron contamination
 in events of the shower counter.
 Filled circles are data between 0.3 and 0.6~GeV/$c$.
 Histograms show the best fit Monte Carlo data.}
\label{fig:efrac}
\end{figure}

The electron contamination was to be considered as a background
 well below 1~GeV/$c$. 
For an estimation of electron contamination,
 events passed through the EM shower section were considered. 
The ${\rm \check{C}erenkov}$ light yield
 from the shower counter was examined to estimate an $e/\mu$ ratio. 
Fig.~\ref{fig:efrac} demonstrates the following estimation procedure. 
The ${\rm \check{C}erenkov}$ light yields
 for electrons and muons were
 individually obtained from Monte Carlo simulations. 
The observed light output distribution is a sum
 of these two distributions with appropriate weights. 
By changing weights for those two distributions,
 the best fit $e/\mu$ ratio was obtained. 
The electron contamination in the muon candidates
 observed in the EM shower section was about 1.5~\% at 0.6~GeV/$c$
 and rapidly decreased with increasing momentum. 
The $e/\mu$ ratio below 1~GeV/$c$ was found to be described
 as $e/\mu = 0.0024 \cdot P^{-3.7}$. 
Above 1~GeV/$c$, the number of observed electron events was too small
 to estimate its ratio by this procedure. 
We estimated the electron contamination
 by extrapolating the ratio obtained below 1~GeV/$c$
 to a higher momentum region. 
After correcting the different data reduction efficiency
 in the normal section from that in the EM shower section
 by using the Monte Carlo simulations,
 the electron contamination was subtracted from the muon candidates
 observed in the normal section. 
Accuracy of the electron subtraction was limited
 by poor statistics of electron events. 
An error in this subtraction was estimated from the fitting procedure
 to be 1.0~\% at 0.6~GeV/$c$
 and rapidly decrease with increasing momentum. 

The proton contamination was to be properly subtracted
 above a few GeV/$c$. 
On the other hand,
 the electron contamination should be
 carefully considered and properly subtracted
 in a different momentum region
 below 1~GeV/$c$. 
Thus the overall error in the contamination subtraction was
 as small as 1~\% over the whole momentum range. 
It may be worth mentioning
 that $e/\mu$ and $p/\mu$ ratios discussed above 
 are different from flux ratios
 because the efficiencies for protons and electrons
 are not same as that for muons. 

\subsection{Corrections}
In order to determine the atmospheric muon spectra
 at the mountain altitude,
 the following corrections are required;
 (i) exposure factor,
 (ii) ionization energy loss, and
 (iii) interaction loss.

The exposure factor is a product
 of geometrical acceptance and live-time. 
The geometrical acceptance defined for this analysis was calculated
 by simulation technique \cite{sullivan1971}
 to be ${\mathrm 0.0214}$~${\mathrm {m^2sr}}$
 for energetic particles which have straight track
 inside the tracking volume. 
It was almost constant over the whole momentum range. 
The simple cylindrical shape and the uniform magnetic field
 make it simple and reliable
 to determine the geometrical acceptance precisely. 
The error arose from uncertainty of the detector alignment
 was estimated to be 0.4~\%. 
Between observed and simulated zenith angular distributions
 of the trajectories,
 a small discrepancy was found below 2~GeV/$c$. 
In the simulation isotropic trajectories were generated
 to obtain the geometrical acceptance,
 however,
 the flux of atmospheric muons has a zenith angle dependence. 
In a low momentum region,
 the dependence might be significant
 even in the so near vertical direction as $\cos \theta _z \geq 0.98$. 
The discrepancy was included in the systematic errors. 
The systematic error associated with this discrepancy was
 estimated to be 3.5~\%
 at 0.6~GeV/$c$ and decreased with increasing momentum. 
The live data-taking time was measured exactly to be 339,874 seconds. 

The energy of each particle at the top of the instrument was calculated
 by summing up the ionization energy losses inside the instrument
 with tracing back the event trajectory. 

In order to estimate the interaction loss probabilities
 inside the BESS detector,
 Monte Carlo simulations
 with the GEANT code \cite{geant} were performed. 
The probabilities
 that muons can be identified as single-track events
 were evaluated
 by applying the same selection criterion to the Monte Carlo events. 
The resultant single-track efficiency was 97.9~\% at 10~GeV/$c$
 and almost constant over the momentum region discussed here. 
The Monte Carlo simulation of the BESS detector
 well reproduced the observed event profile. 
By comparing the observed and simulated parameters
 used in the single-track selection, i.e.,
 the number of tracks inside the tracking volume
 and the number of hit-counters in each layer of the TOF hodoscopes,
 it was concluded that
 a discrepancy between the observed and simulated event shapes was
 no more than $\pm 1.6\%$ in the whole momentum range discussed here. 
This discrepancy was included in the systematic error
 of the single-track efficiency. 
Another source of the systematic error in the Monte Carlo simulation was
 uncertainty of the material distribution inside the BESS detector,
 which was estimated to be $\pm 10$~\%. 
Since the single-track efficiency was as high as 98~\%,
 the systematic error originated from this uncertainty
 was as small as 0.2~\%. 

The combined systematic errors
 obtained by quadratically adding the uncertainties
 were $\pm 3.0$, $\pm 1.4$, and $\pm 1.7$~\%
 at 1, 10, and 100~GeV/c, respectively. 

\subsection{Spectrum deformation effect}
Because of the finite resolution in the rigidity measurement
 and the steep spectral shape,
 the observed spectrum may suffer deformation. 
In a low rigidity region well below the MDR,
 the spectrum deformation is negligibly small. 
In a higher rigidity region, however,
 a deflection becomes comparable to an error in curvature measurement;
 thus, the spectrum deformation is to be considered. 
This spectrum deformation effect of the BESS detector was discussed
 in our previous paper \cite{bessphe}. 
According to the discussion,
 the effect was estimated to be no more than 3~\% below 120~GeV/$c$. 
The deformation effect is smaller than the statistical errors
 over the whole momentum range in this analysis. 
Thus no correction was applied to the measured spectra. 

\section{Results and discussions} \label{sec:result}

\begin{figure}
\includegraphics[width=\textwidth]{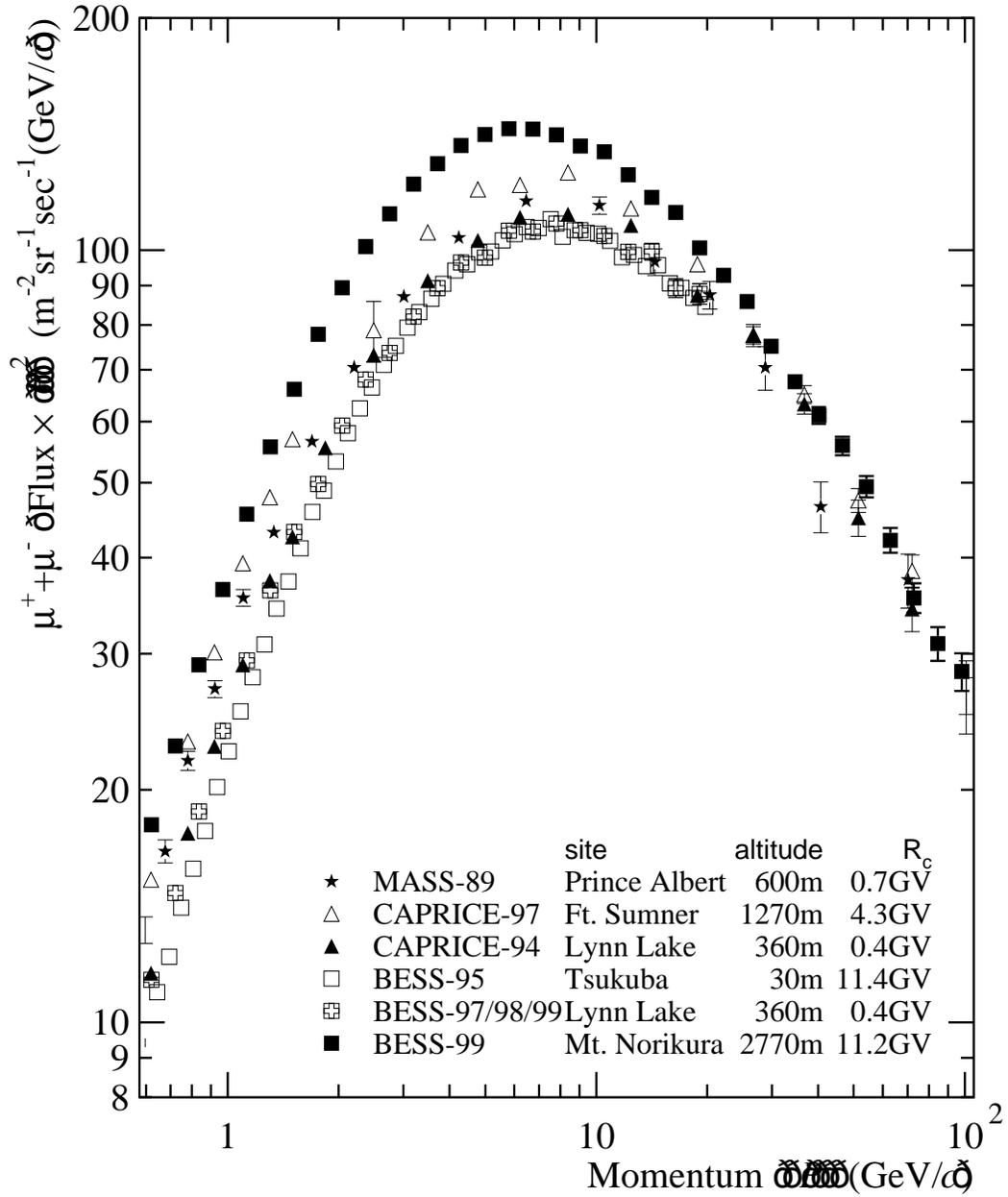}
\caption{Absolute differential momentum spectra of atmospheric muons.
 Vertical axis shows the sum of $\mu^+$ and  $\mu^-$ fluxes
 multiplied by $P^2$.
 Only statistical errors are included.
 The geomagnetic cutoff rigidity at each site is shown
 as $R_{\mathrm{c}}$.
 The difference among the fluxes is
 mainly due to the difference in altitudes. }
\label{fig:muf}
\end{figure}

The momentum spectra of the atmospheric muons
 in momentum ranges of 0.6 through 106 GeV/$c$
 at a mountain altitude of 2,770~m above sea level
 have been measured with the BESS detector. 
The results are summarized in Table~\ref{tbl:result}
 and shown in Fig.~\ref{fig:muf}. 
The first and second errors in Table~\ref{tbl:result}
 represent statistical and systematic errors, respectively. 
The overall errors including both errors
 are less than $\pm 10$~\%. 
In Fig.~\ref{fig:muf},
 our result is compared
 with other absolute flux measurements \cite{motokimu,pascal,kremer}. 
A discrepancy between the observed muon fluxes is
 mainly due to the difference in altitudes. 

\begin{figure}
\includegraphics[width=\textwidth]{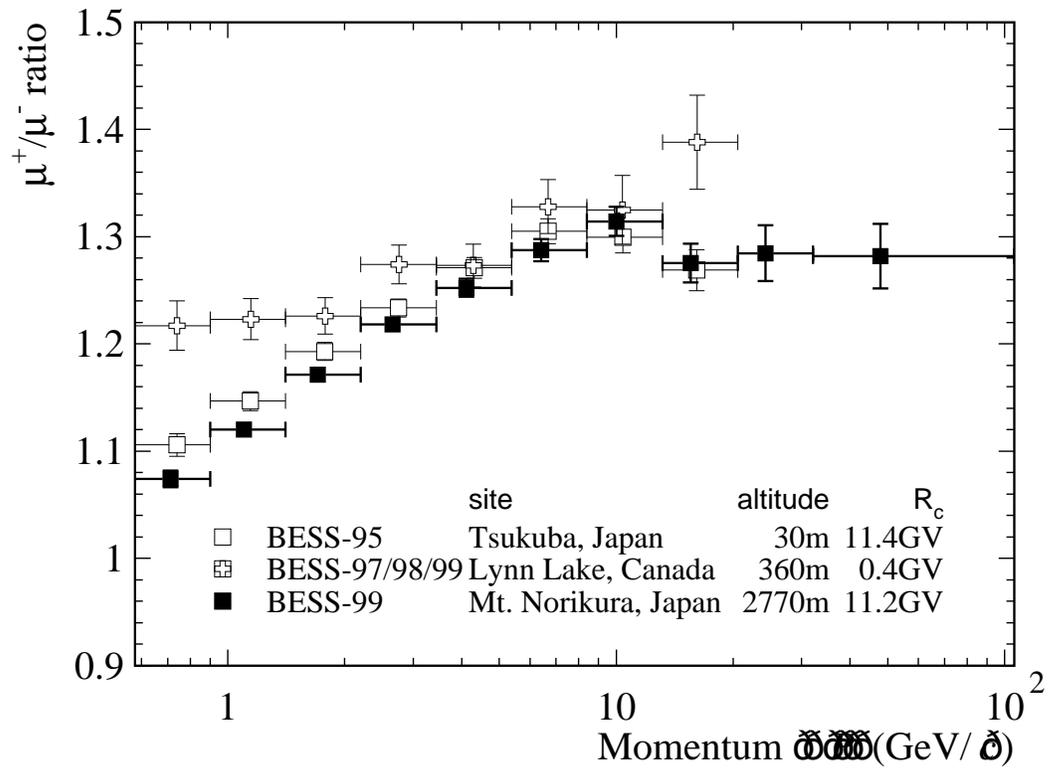}
\caption{Charge ratios of atmospheric muons.
 Only statistical errors are included.
 The geomagnetic cutoff rigidity at each site is shown
 as $R_{\mathrm{c}}$.
 The difference among the ratios is caused
 by the difference in geomagnetic cutoff rigidity
 and energy loss inside the atmosphere. }
\label{fig:ratio}
\end{figure}

Fig.~\ref{fig:ratio} shows charge ratios of atmospheric muons
 observed in a series of muon measurements with the BESS detector. 
The big difference in the charge ratios observed in Japan and Canada
 comes from the influence of the geomagnetic cutoff rigidity
 as discussed in Ref.~\cite{motokimu}. 
Below 10~GeV/$c$, the charge ratio at mountain altitude
 is slightly lower than that at sea level. 
This tendency was made by energy loss inside the atmosphere
 between mountain altitude and sea-level. 
In a higher momentum region,
 the ratio seems to approach to around 1.28. 

\begin{figure}
\includegraphics[width=\textwidth]{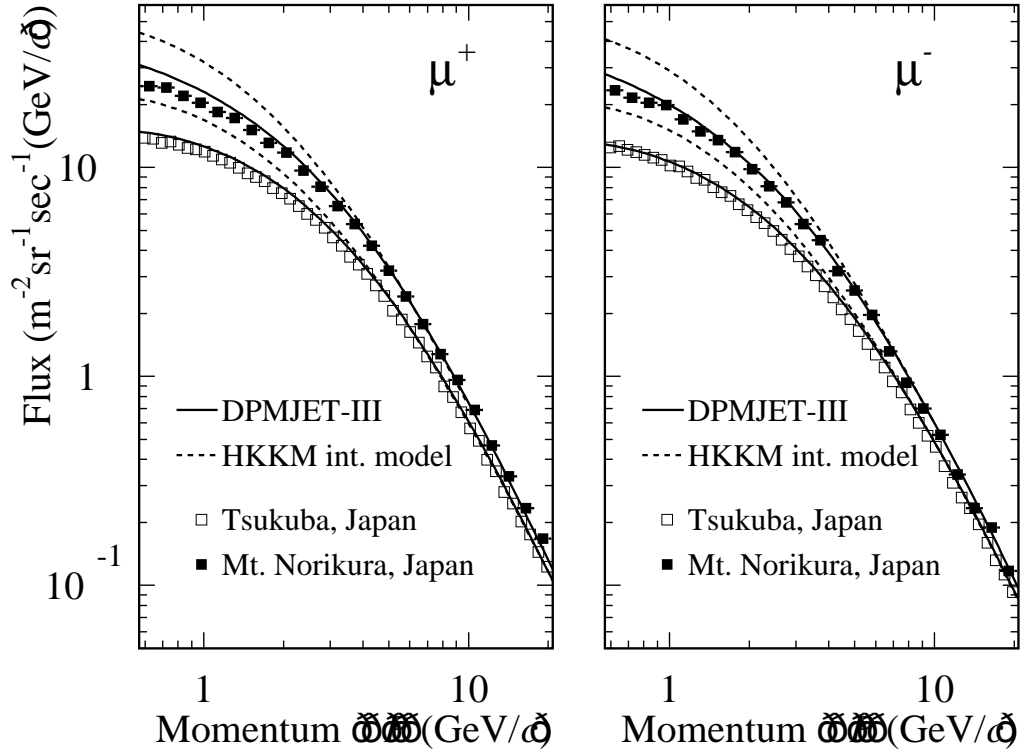}
\caption{Comparison between the observed and
 calculated spectra of atmospheric muons.
 The calculation with the DPMJET-III event generator
 reproduced our results better than the HKKM interaction model.}
\label{fig:japan-mu}
\end{figure}

Our results were compared with theoretical predictions,
 in which atmospheric muon spectra were calculated
 simultaneously with atmospheric neutrinos. 
Fig.~\ref{fig:japan-mu} shows comparisons
 between our results and the predictions. 
The predicted spectra were calculated based on Ref.~\cite{hkkm2001},
 but some parameters,
 such as atmospheric pressure, temperature and zenith angle limit,
 were adjusted to reproduce the experimental conditions. 
In Ref.~\cite{hkkm2001},
 two calculations were discussed;
 one was developed
 on the basis of the HKKM interaction model \cite{hkkm1995}
 and the other was calculated
 by employing the DPMJET-III event generator \cite{dpmjet3}
 as hadronic interaction models. 
As Fig.~\ref{fig:japan-mu} indicates,
 the DPMJET-III interaction model
 reproduced the observed spectra
 better than the HKKM interaction model. 
Some deviation, however, was found
 between the observed and calculated spectra
 especially below 1~GeV/$c$. 
It seems reasonable to suppose
 that the hadronic interaction models are to be tested and improved
 by accurate spectra of atmospheric muons. 
The precise measurement of atmospheric secondary particles
 will improve the accuracy in the atmospheric neutrino calculations. 

\section{Summary} \label{sec:aummary}

We have measured absolute fluxes of atmospheric muons
 at the top of Mt. Norikura, Japan.
It is located at an altitude of 2,770~m above sea level. 
The overall errors of our measurements are less than $\pm 10$~\%. 
The absolute fluxes showed relatively good agreement
 with theoretical predictions
 calculated by using the DPMJET-III hadronic interaction model. 
The precise measurement of the atmospheric muons
 can modify the hadronic interaction model
 and improve accuracy of atmospheric neutrino calculations.

\begin{ack}
We would like to thank all staffs in the Norikura Observatory, ICRR,
 the University of Tokyo for their cooperation. 
We are indebted to M. Honda and K. Kasahara
 for their Monte Carlo simulations and theoretical interpretations. 
This experiment was supported by Grants-in-Aid
 from the ministry of Education, Culture, Sport, Science and Technology,
 MEXT. 
We would like to thank ISAS and KEK for their continuous support
 and encouragement for the BESS experiment. 
The analysis was performed with the computing facilities
 at ICEPP, the University of Tokyo. 
\end{ack}

\clearpage
%
%

\clearpage

\clearpage

\clearpage

\clearpage

\clearpage

\clearpage

\clearpage

\clearpage

\clearpage
%
%
\begin{table}
\caption{Absolute differential momentum spectra of atmospheric muons.}
\label{tbl:result}
\end{table}\begin{supertabular}{ccclccl}
\hline
 Momentum 
& & \multicolumn{2}{c}{$\mu ^+$} & & \multicolumn{2}{c}{$\mu ^-$} \\
    \cline{3-4}                      \cline{6-7}
     range     & & 
$\overline{P}$ & ${\rm Flux\pm\Delta Flux_{sta}\pm\Delta Flux_{sys}}$ &&
$\overline{P}$ & ${\rm Flux\pm\Delta Flux_{sta}\pm\Delta Flux_{sys}}$\\ 
      GeV/$c$  & &
GeV/$c$& \multicolumn{1}{c}{$({\rm m^{2}~sr~sec~GeV/}c)^{-1}$} & &
GeV/$c$&\multicolumn{1}{c}{$({\rm m^{2}~sr~sec~GeV/}c)^{-1}$} \\
\hline
  .576 -- .669 & & .623 & 2.39 $\pm$ 0.02 $\pm$ 0.09 $\times 10$      & & .623 & 2.29 $\pm$ 0.02 $\pm$ 0.09 $\times 10$      \\ 
  .669 -- .776 & & .722 & 2.29 $\pm$ 0.02 $\pm$ 0.08 $\times 10$      & & .723 & 2.11 $\pm$ 0.02 $\pm$ 0.07 $\times 10$      \\
  .776 -- .901 & & .839 & 2.17 $\pm$ 0.02 $\pm$ 0.07 $\times 10$      & & .838 & 1.99 $\pm$ 0.02 $\pm$ 0.06 $\times 10$      \\
  .901 -- 1.05 & & .972 & 2.03 $\pm$ 0.02 $\pm$ 0.06 $\times 10$      & & .972 & 1.83 $\pm$ 0.01 $\pm$ 0.05 $\times 10$      \\
  1.05 -- 1.21 & & 1.13 & 1.89 $\pm$ 0.01 $\pm$ 0.05 $\times 10$      & & 1.13 & 1.70 $\pm$ 0.01 $\pm$ 0.05 $\times 10$      \\
  1.21 -- 1.41 & & 1.31 & 1.73 $\pm$ 0.01 $\pm$ 0.05 $\times 10$      & & 1.31 & 1.53 $\pm$ 0.01 $\pm$ 0.04 $\times 10$      \\
  1.41 -- 1.63 & & 1.52 & 1.54 $\pm$ 0.01 $\pm$ 0.04 $\times 10$      & & 1.52 & 1.33 $\pm$ 0.01 $\pm$ 0.03 $\times 10$      \\
  1.63 -- 1.90 & & 1.76 & 1.36 $\pm$ 0.01 $\pm$ 0.03 $\times 10$      & & 1.76 & 1.16 $\pm$ 0.01 $\pm$ 0.03 $\times 10$      \\
  1.90 -- 2.20 & & 2.05 & 1.16 $\pm$ 0.01 $\pm$ 0.03 $\times 10$      & & 2.04 & 9.79 $\pm$ 0.07 $\pm$ 0.22                  \\ 
  2.20 -- 2.55 & & 2.37 & 9.84 $\pm$ 0.06 $\pm$ 0.21                  & & 2.37 & 8.13 $\pm$ 0.06 $\pm$ 0.17                  \\
  2.55 -- 2.96 & & 2.75 & 8.09 $\pm$ 0.05 $\pm$ 0.16                  & & 2.75 & 6.64 $\pm$ 0.05 $\pm$ 0.13                  \\
  2.96 -- 3.44 & & 3.19 & 6.57 $\pm$ 0.04 $\pm$ 0.13                  & & 3.19 & 5.36 $\pm$ 0.04 $\pm$ 0.11                  \\
  3.44 -- 3.99 & & 3.71 & 5.21 $\pm$ 0.04 $\pm$ 0.10                  & & 3.71 & 4.21 $\pm$ 0.03 $\pm$ 0.08                  \\
  3.99 -- 4.63 & & 4.30 & 4.14 $\pm$ 0.03 $\pm$ 0.08                  & & 4.30 & 3.24 $\pm$ 0.03 $\pm$ 0.06                  \\
  4.63 -- 5.38 & & 4.99 & 3.14 $\pm$ 0.02 $\pm$ 0.06                  & & 4.99 & 2.53 $\pm$ 0.02 $\pm$ 0.05                  \\
  5.38 -- 6.24 & & 5.79 & 2.42 $\pm$ 0.02 $\pm$ 0.04                  & & 5.79 & 1.87 $\pm$ 0.02 $\pm$ 0.03                  \\
  6.24 -- 7.25 & & 6.72 & 1.78 $\pm$ 0.02 $\pm$ 0.03                  & & 6.71 & 1.39 $\pm$ 0.01 $\pm$ 0.02                  \\
  7.25 -- 8.41 & & 7.80 & 1.30 $\pm$ 0.01 $\pm$ 0.02                  & & 7.80 & 1.01 $\pm$ 0.01 $\pm$ 0.02                  \\
  8.41 -- 9.76 & & 9.05 & 9.48 $\pm$ 0.10 $\pm$ 0.16 $\times 10^{-1}$ & & 9.05 & 7.16 $\pm$ 0.09 $\pm$ 0.12 $\times 10^{-1}$ \\
  9.76 -- 11.3 & & 10.5 & 6.83 $\pm$ 0.08 $\pm$ 0.19 $\times 10^{-1}$ & & 10.5 & 5.31 $\pm$ 0.07 $\pm$ 0.15 $\times 10^{-1}$ \\
  11.3 -- 13.1 & & 12.2 & 4.81 $\pm$ 0.06 $\pm$ 0.13 $\times 10^{-1}$ & & 12.2 & 3.60 $\pm$ 0.05 $\pm$ 0.10 $\times 10^{-1}$ \\
  13.1 -- 15.3 & & 14.1 & 3.29 $\pm$ 0.05 $\pm$ 0.09 $\times 10^{-1}$ & & 14.2 & 2.55 $\pm$ 0.04 $\pm$ 0.07 $\times 10^{-1}$ \\
  15.3 -- 17.7 & & 16.4 & 2.31 $\pm$ 0.04 $\pm$ 0.06 $\times 10^{-1}$ & & 16.4 & 1.83 $\pm$ 0.03 $\pm$ 0.05 $\times 10^{-1}$ \\
  17.7 -- 20.6 & & 19.0 & 1.55 $\pm$ 0.03 $\pm$ 0.04 $\times 10^{-1}$ & & 19.0 & 1.22 $\pm$ 0.02 $\pm$ 0.03 $\times 10^{-1}$ \\
  20.6 -- 23.9 & & 22.1 & 1.07 $\pm$ 0.02 $\pm$ 0.03 $\times 10^{-1}$ & & 22.1 & 8.25 $\pm$ 0.19 $\pm$ 0.23 $\times 10^{-2}$ \\
  23.9 -- 27.7 & & 25.6 & 7.21 $\pm$ 0.16 $\pm$ 0.20 $\times 10^{-2}$ & & 25.6 & 5.79 $\pm$ 0.15 $\pm$ 0.16 $\times 10^{-2}$ \\
  27.7 -- 32.1 & & 29.8 & 4.81 $\pm$ 0.12 $\pm$ 0.13 $\times 10^{-2}$ & & 29.8 & 3.63 $\pm$ 0.11 $\pm$ 0.10 $\times 10^{-2}$ \\
  32.1 -- 37.3 & & 34.5 & 3.18 $\pm$ 0.09 $\pm$ 0.09 $\times 10^{-2}$ & & 34.6 & 2.46 $\pm$ 0.08 $\pm$ 0.07 $\times 10^{-2}$ \\
  37.3 -- 43.3 & & 39.9 & 2.10 $\pm$ 0.07 $\pm$ 0.06 $\times 10^{-2}$ & & 40.1 & 1.69 $\pm$ 0.06 $\pm$ 0.05 $\times 10^{-2}$ \\
  43.3 -- 50.2 & & 46.5 & 1.45 $\pm$ 0.05 $\pm$ 0.04 $\times 10^{-2}$ & & 46.6 & 1.12 $\pm$ 0.05 $\pm$ 0.03 $\times 10^{-2}$ \\
  50.2 -- 58.3 & & 54.1 & 9.85 $\pm$ 0.41 $\pm$ 0.28 $\times 10^{-3}$ & & 53.9 & 7.03 $\pm$ 0.35 $\pm$ 0.20 $\times 10^{-3}$ \\
  58.3 -- 67.7 & & 62.7 & 5.85 $\pm$ 0.29 $\pm$ 0.17 $\times 10^{-3}$ & & 62.6 & 4.83 $\pm$ 0.27 $\pm$ 0.14 $\times 10^{-3}$ \\
  67.7 -- 78.5 & & 73.2 & 3.77 $\pm$ 0.22 $\pm$ 0.11 $\times 10^{-3}$ & & 72.3 & 2.90 $\pm$ 0.19 $\pm$ 0.08 $\times 10^{-3}$ \\
  78.5 -- 91.1 & & 84.5 & 2.29 $\pm$ 0.16 $\pm$ 0.07 $\times 10^{-3}$ & & 84.6 & 2.03 $\pm$ 0.15 $\pm$ 0.06 $\times 10^{-3}$ \\
  91.1 -- 106. & & 97.4 & 1.70 $\pm$ 0.13 $\pm$ 0.05 $\times 10^{-3}$ & & 97.3 & 1.25 $\pm$ 0.11 $\pm$ 0.04 $\times 10^{-3}$ \\
\hline
\end{supertabular}


\begin{thebibliography}{00}

\bibitem{bessphe}
  T. Sanuki et al.,
  ApJ 545 (2000) 1135.

\bibitem{amsp}
  AMS Collab., J. Alcaraz et al.,
  Phys. Lett. B 490 (2000) 27.

\bibitem{amshe}
  AMS Collab., J. Alcaraz et al.,
  Phys. Lett. B 494 (2000) 193.

\bibitem{motokimu}
  M. Motoki et al.,
  astro-ph/0205344 (2002), submitted to Astropart. Phys.

\bibitem{orito}
  S. Orito,
  in : Proc. ASTROMAG Workshop, KEK Report KEK87-19,
  eds. J. Nishimura, K. Nakamura, and A. Yamamoto
  (KEK, Ibaraki, 1987)p.111.

\bibitem{yamamoto1994}
  A. Yamamoto et al.,
  Adv. Space Res. 14 (1994) 75.

\bibitem{agel}
  Y. Asaoka et al.,
  Nucl. Instr. and Meth. A 416 (1998) 236.

\bibitem{detector}
  Y. Ajima et al.,
  Nucl. Instr. and Meth. A 443 (2000) 71.

\bibitem{newtof}
  Y. Shikaze et al.,
  Nucl. Instr. and Meth. A 455 (2000) 596.

\bibitem{shea}
  M. A. Shea and D. F. Smart,
  Proc. 27th ICRC(Hamburg) (2001) 4063.

\bibitem{cosmos}
  K. Kasahara.,
  Proc. 24th ICRC(Rome) 1 (1995) 399, also, \\
  http://eweb.b6.kanagawa-u.ac.jp/\verb|~|kasahara/ResearchHome/cosmosHome/index.html.

\bibitem{fritiof7}
  H. Pi,
  Comput. Phys. Commun. 71 (1992) 173.

\bibitem{dpmjet3}
  S. Roesler, R. Engel, and J. Ranft,
  SLAC-PUB-8740, hep-ph/0012252 (2000), unpublished.

\bibitem{sullivan1971}
  J. D. Sullivan, 
  Nucl. Instr. and Meth. 95 (1971) 5.

\bibitem{geant}
  R. Brun et al.,
  GEANT -- Detector Description and Simulation Tool,
  CERN program library
  (CERN, Geneva, 1994).

\bibitem{pascal}
  M. P. De Pascal et al.,
  J. Geophys. Res. 98 (1993) 3501.

\bibitem{kremer}
  J. Kremer et al.,
  Phys. Rev. Lett. 83 (1999) 4241.

\bibitem{hkkm2001}
  M. Honda et al.,
  Proc. 27th ICRC(Hamburg) (2001) 1162.

\bibitem{hkkm1995}
  M. Honda et al.,
  Phys. Rev. D 52 (1995) 4985.

\end{thebibliography}
\end{document}